# Advanced iterative procedures for solving the implicit Colebrook equation for fluid flow friction

Pavel Praks[1,2,*], Dejan Brkić[1,*]

[1]European Commission, DG Joint Research Centre (JRC), Directorate C: Energy, Transport and Climate, Unit C3: Energy Security, Distribution and Markets, Via Enrico Fermi 2749, 21027 Ispra (VA), Italy

[2] IT4Innovations National Supercomputing Center, VŠB - Technical University Ostrava, 17. listopadu 2172/15, 708 00 Ostrava, Czech Republic

pavel.praks@ec.europa.eu (Pavel Praks) - ORCID id: https://orcid.org/0000-0002-3913-7800
dejanbrkic0611@gmail.com (Dejan Brkić) - ORCID id: https://orcid.org/0000-0002-2502-0601

*both authors contributed equally to this study

**Abstract:** Empirical Colebrook equation from 1939 is still accepted as an informal standard to calculate friction factor during the turbulent flow ($4000<Re<10^8$) through pipes from smooth with almost negligible relative roughness ($\varepsilon/D \rightarrow 0$) to the very rough (up to $\varepsilon/D=0.05$) inner surface. The Colebrook equation contains flow friction factor $\lambda$ in implicit logarithmic form where it is, aside of itself; $\lambda$, a function of the Reynolds number Re and the relative roughness of inner pipe surface $\varepsilon/D$; $\lambda=f(\lambda, Re, \varepsilon/D)$. To evaluate the error introduced by many available explicit approximations to the Colebrook equation, $\lambda \approx f(Re, \varepsilon/D)$, it is necessary to determinate value of the friction factor $\lambda$ from the Colebrook equation as accurate as possible. The most accurate way to achieve that is using some kind of iterative methods. Usually classical approach also known as simple fixed point method requires up to 8 iterations to achieve the high level of accuracy, but does not require derivatives of the Colebrook function as here presented accelerated Householder's approach ($3^{rd}$ order, $2^{nd}$ order: Halley's and Schröder's method and $1^{st}$ order: Newton-Raphson) which needs only 3 to 7 iteration and three-point iterative methods which needs only 1 to 4 iteration to achieve the same high level of accuracy. Strategies how to find derivatives of the Colebrook function in symbolic form, how to avoid use of the derivatives (Secant method) and how to choose optimal starting point for the iterative procedure are shown. Householder's approach to the Colebrook's equations expressed through the Lambert W-function is also analyzed. One approximation to the Colebrook equation based on the analysis from the paper with the error of no more than 0.0617% is shown.

## 1. Introduction

To evaluate flow resistance in turbulent flow through pipes, from smooth to rough, the empirical Colebrook equation is in common use (1) {Colebrook 1939}:

$$\underbrace{\frac{1}{\sqrt{\lambda}} = -2 \cdot \log_{10}\left(\frac{2.51}{Re \cdot \sqrt{\lambda}} + \frac{\varepsilon}{3.7 \cdot D}\right)}_{Colebrook} \tag{1}$$

In the Colebrook equation $\lambda$ represents Darcy flow friction factor, Re Reynolds number and $\varepsilon/D$ relative roughness of inner pipe surfaces (all three quantities are dimensionless).

The experiment performed by Colebrook and White {Colebrook and White 1937} dealt with flow of air through one pipe (diameter D=53.5mm and length L=6m) with six different roughness of inner surface of the pipe artificially simulated with various mixtures of two sizes of sand grain (0.035mm and 0.35mm diameter) to simulate conditions of inner pipe surface from almost smooth to very rough. The sand grains were fixed using a sort of bituminous adhesive waterproof insulating compound to form five types of relatively uniform roughness of inner pipe surfaces while the sixth one was without sand, i.e. it was left smooth. The experiment revealed, contrary to the previous, that flow friction, $\lambda$ does not have a sharp transition from the smooth to the fully rough law of turbulence. This evidence Colebrook {Colebrook 1939} later captured in the today famous and widely used empirical equation; Eq. (1).

The Colebrook function relates the unknown flow friction factor $\lambda$ as function of itself, the Reynolds number Re and the relative roughness of inner pipe surface $\varepsilon/D$; $\lambda=f(\lambda, Re, \varepsilon/D)$. It is valid for $4000<Re<10^8$ and for $0<\varepsilon/D<0.05$. The Colebrook equation is transcendent and thus cannot be solved in terms of elementary functions {Sonnad and Goudar 2007, Brkić 2011a, Mikata and Walczak 2016, Vatankhah 2018}. Although empirical, and therefore with questionable accuracy its precise solution is sometimes essential in order to repeat or to evaluate previous findings in a concise way {Schockling et al. 2007, Allen et al. 2007, Langelandsvik et al. 2008, Clamond 2009}.

Few approaches are available today for solving Colebrook equation:

(1) **Graphical solution – Moody diagram:** To represent the Colebrook equation graphically, Rouse in 1942 had developed appropriate diagram which Moody later adopted in 1944 in famous diagram widely used in the past in engineering practice {Moody 1944, LaViolette 2017}. The diagram was preferred because the Colebrook equation is implicitly given. Today, graphical solution has only value for educational purposes.
(2) **Iterative solution of the Colebrook equation:**
*(2.a.) Simple fixed point iterative method.* The simple fixed point iterative method {Brkić 2017a} is in common use for solving accurately the Colebrook equation (special case of the Colebrook equation for Re→∞ gives explicit form valid only for the fully turbulent flow in rough pipes {Schultz and Flack 2007, Herwig et al. 2008, Brkić 2012a, Brkić 2016a} but which can be used as initial starting point for all cases covered by the Colebrook equation $\lambda_0=f(\varepsilon/D)\rightarrow$Eq. (2); now using the Colebrook equation new value can be calculated $\lambda_1=f(\lambda_0; Re; \varepsilon/D)$; starting from i=1, the procedure $\lambda_{i+1}=f(\lambda_i; Re; \varepsilon/D)$; i=i+1 goes until $\lambda_i \approx \lambda_{i+1}$, where we set $\lambda_{i+1}-\lambda_i \leq 10^{-8}$). It usually reaches the satisfied accuracy after about 8 iterations.

*(2.b.) Householder's iterative methods.* On the other hand, Newton's method (also known as the Newton-Raphson method {Cajori 1911, Ypma 1995, Abbasbandy 2003}) needs only 3 to 7 iterations to reach the same level of accuracy. A shortcoming of Newton's method is that it additionally requires the first derivative of the Colebrook's function (here we show analytical form of the first derivative including the symbolic form generated in MATLAB {Shampine 2007}). Also knowing that the Newton-Raphson method is $1^{st}$ order of Householder's method {Householder 1970, Griffiths and Smith 2006}, we analyze $2^{nd}$ order which is known as Halley's {Halley 1694} and Schröder's {Schröder 1870, Petković et al. 2010} method, and also $3^{rd}$ order. The $3^{rd}$ order methods use the third, the second and the first derivative, $2^{nd}$ order the second and the first while $1^{st}$ order use only the first derivative. Today, all mentioned types of iterative solutions can easily be implemented in software codes and they has been accepted as the most accurate way for solving the Colebrook equation and hence they are preferred compared to the graphical solution.

*(2.c.) Three-point iterative methods.* Three-point iterative methods needs only 1 iteration in three points $x_0$, $y_0$ and $z_0$ (three internal iterations) to achieve high level of accuracy {Džunić et al. 2011, Petković et al. 2014, Sharma and Arora 2016,}. $x_0$ is initial starting point, $y_0$ is auxiliary step, while $z_0$ is the solution. Three-point methods are very accurate and can reach high accuracy some cases even after 1 to 2 iterations. Also slightly less accurate two-point methods exist.

**(3) Approximations of the Colebrook equation:** Colebrook's equation can be expressed in explicit form only in an approximate way; $\lambda \approx f(Re, \varepsilon/D)$ {Gregory and Fogarasi 1985, Zigrang and Sylvester 1985, Giustolisi et al. 2011, Genić et al. 2011, Brkić 2011b, Brkić 2012b, Winning and Coole 2013, Brkić and Ćojbašić 2017}. Numerous explicit approximations to the Colebrook equation are available in literature {Gregory and Fogarasi 1985, Zigrang and Sylvester 1985, Brkić 2011b, Genić et al. 2011, Brkić 2012b, Winning and Coole 2013, Brkić and Ćojbašić 2017}. Iterative solution as the most accurate method is used for evaluation of accuracy of such approximations. Also, based on our findings, we provide an approximation; Eq. (28) with the error of no more than 0.69% and 0.0617%. The Colebrook equation also can be approximately simulated using Artificial Neural Networks {Özger and Yıldırım 2009, Brkić and Ćojbašić 2016, Bardestani et al. 2017}.

**(4) Lambert W-function:** Until now only one known way to express the Colebrook equation exactly in explicit way is through the Lambert W-function; $\lambda=W(Re, \varepsilon/D)$ {Keady 1998, Clamond 2009, Brkić 2011a, Brkić 2012c}, where further evaluation of the Lambert W-function can be only approximate {Corless et al. 1996, Boyd 1998, Barry et al. 2000, Hayes 2005, Mező and Baricz 2017}. Here we show procedure how to solve the Lambert W-function using Householder's iterative procedure ($2^{nd}$ order: Halley's method and $1^{st}$ order: Newton-Raphson). Also approach with the shifted Lammert W-function in term of the Wright Ω-function exist {Rollmann and Spindler 2015, Biberg 2017}.

In this paper, we show Householder's iterative procedures (3$^{rd}$ order, 2$^{nd}$ order: Halley's {Scavo and Thoo 1995} and Schröder's method and 1$^{st}$ order: Newton-Raphson) and one example of the three-point methods with additional recommendations in order to solve the empirical Colebrook equation which is implicitly given in respect to flow friction factor λ. The goal of this paper is to show improved iterative solutions that can obtain value of the unknown friction factor λ accurately after the least possible number of iterations. Additionally we developed a strategy how to choose the best starting point {Kornerup and Muller 2006} for the iterative procedure in the domain of interest of the Colebrook equation, how to generate required symbolic derivatives to the Colebrook equation in MATLAB and how to avoid use of derivatives (Secant method). Finally, we use findings from our paper to present a novel explicit approximation of the Colebrook equation, which would be interesting for engineering practice. We also present distribution of the relative error in respect of the presented approximation over the applicability domain of the Colebrook equation.

To evaluate efficiency of the presented methods unknown flow friction factor λ is calculated for two pairs of the Reynolds number Re and relative roughness of inner pipe surfaces ε/D; 1) {Re=5·10$^6$, ε/D=2.5·10$^{-5}$}→λ=0.010279663295529, and 2) {Re=3·10$^4$, ε/D=9·10$^{-3}$}→λ=0.038630738574792.

## 2. Initial estimate of starting point for the iterative procedures

The starting point is a significant factor in convergence speed in three-point and Householder's method {Kornerup and Muller 2006} and there are different methods to choose a good start but here we examine 1) Starting point as function of input parameters, and 2) Initial starting point with the fixed value.

One of the essential issues in every iterative procedure is to choose good starting point {Moursund 1967, Taylor 1970}. Here we try to find the fixed starting point (initial value of the flow friction factor $\lambda_0$ or the related transmission factor $x_0 = \frac{1}{\sqrt{\lambda_0}}$) valid for all cases from the practical domains of applicability of the Colebrook equation which is; Reynolds number Re; 4000<Re<10$^8$ and the relative roughness ε/D; 0<ε/D<0.05. In the cases when this approach does not work efficiently we show how to choose the starting value in function of the Reynolds number Re and the relative roughness ε/D, i.e. using some kind of rough approximation to the Colebrook equation which can be relatively inaccurate but simply and which put the initial value reasonable close to the final and accurate solution. This initial guess then need to be plugged into the shown numerical methods and iterated recursively few times (usually two or three times) to converge upon the final solution. In any case, a sample of size 65536 was considered for analysis of iteration methods. The input sample was generated according to the uniform density function of each input variable. The low-discrepancy Sobol sequences were employed {Sobol et al. 1992}. These so called quasirandom sequences have useful properties. In contrary to random numbers, quasirandom numbers covers the space more quickly and evenly. Thus, they leave very few holes.

The Colebrook equation also can be expressed in term of the Lambert W-function analytically; λ=f(λ, Re, ε/D)→λ=W(Re, ε/D) {Keady 1998, Sonnad and Goudar 2004, Brkić 2012c}. The Lambert W-function further can be evaluated only approximately through Householder's iterative procedures which also require the appropriate initial starting point. The analysis of this initial starting point has wider applicability, because the Lambert W-function has extensive use in many branches of physics and technology {Valluri et al. 2000, Hosseini et al. 2014}.

### 2.1. Starting point as function of input parameters

### 2.1.1. Starting point as function of the relative roughness ε/D (when Re→∞)

Special case of the Colebrook equation when Re→∞ physically means that the flow friction factor λ in that case depends only on ε/D; for Re→∞, λ=f(ε/D), i.e. the flow friction factor λ is not implicitly given {Brkić 2016a}. In that way starting point can be calculated using explicit equation which has only one variable; $\lambda_0$=f(ε/D); Eq. (2). The results obtained in that way are accurate only for the case Re→∞ but for the smaller values of Re which corresponds to the smooth turbulent flow the error can goes up to 80% {Brkić 2011c, Brkić 2012a}. Anyway, in that way calculated value can be efficiently used as initial starting guess for iterative procedure for the whole domain of applicability of the Colebrook equation.

$$x_0 = \frac{1}{\sqrt{\lambda_0}} = \underbrace{-2 \cdot \log_{10}\left(\frac{\varepsilon}{3.7 \cdot D}\right)}_{rough\ part\ of\ Colebrook} \qquad (2)$$

The initial starting point obtained using the previous equation is referred as "traditional", it introduces the maximal relative error of 80% over the domain of applicability of the Colebrook equation where the error can be neglected in case of fully developed turbulent flow through the pipes with very rough inner surface. To reach accuracy of $\lambda_{i+1}-\lambda_i \leq 10^{-8}$ usually 6 steps are enough regarding the Newton-Raphson method (Figure 1).

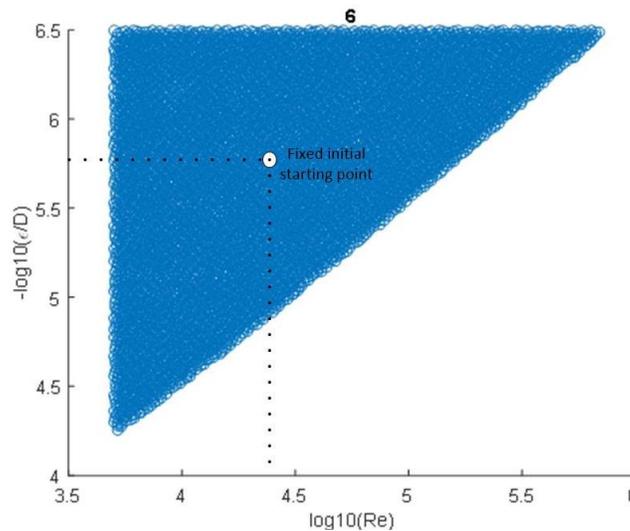

Figure 1. "Slow area" which requires 6 iterations to reach accuracy of $10^{-8}$ regarding the "traditional" option for the starting point calculated through Eq. (2), for solving Colebrook's equation in Newton's procedure when calculation goes through the transmission factor x

**2.1.2 Starting point obtained using approximations to the Colebrook equation**

Every approximation to the Colebrook equation; λ≈f(Re, ε/D) can be used to put initial starting point as close as possible near the final accurate solution {Brkić 2011b}. For example, using one of the approximation with the error of up to 10% for calculation of initial starting point $x_0 = \frac{1}{\sqrt{\lambda}}$, the final accurate value of the flow friction factor λ is reached in the worst case scenario after 3 iterations using Colebrook's equation and one of the procedures from Section 3.2. After 3 iterations, the whole practical domain of applicability of the Colebrook equation is covered with the difference between the two final iterations less than $10^{-8}$; $\lambda_{i+1}-\lambda_i \leq 10^{-8}$ (Figure 2). In average, the method requires 2.7 iterations in average for all cases with set precision (stopping criterion) very close to zero (about $10^{-8}$) when calculation goes through the transmission factor x. The results from Figure 1 are from the 65536 pairs of the Reynolds number Re and the relative roughness ε/D over the domain of applicability of the Colebrook equation domain (values of the Reynolds number Re between 4000 and $10^8$ and the relative roughness ε/D between 0 and 0.05, dividing them into 256 points each).

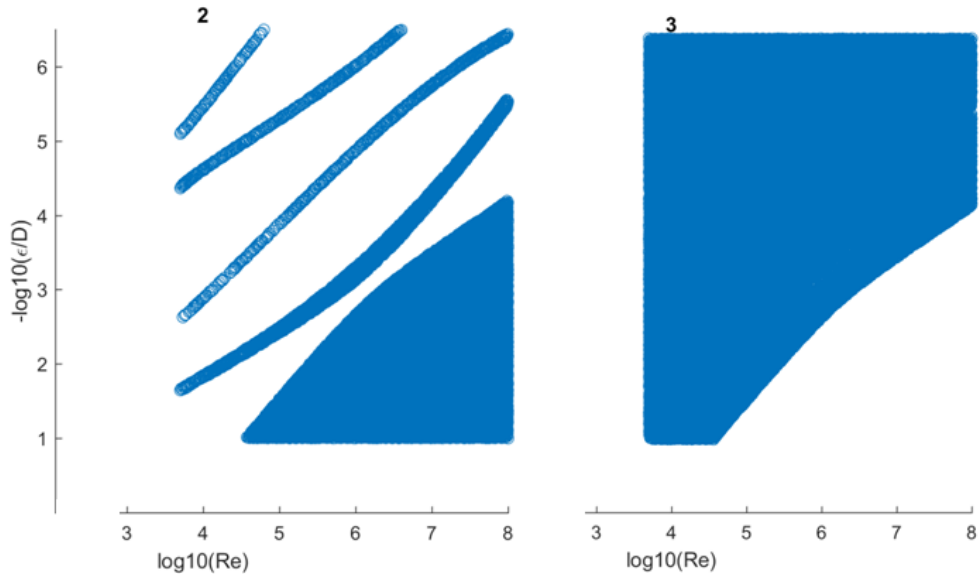

Figure 2. Area in which 2 iterations (left), and 3 iterations (right) are sufficient to calculate the flow friction factor λ with accuracy of $10^{-8}$ using Colebrook equation solved in Newton's procedure when calculation goes through the transmission factor x and using approximation with error of up to 10%

**2.2 Fixed initial starting point**

An idea from geometry to find "center of gravity" is used for the points for which the Newton-Raphson, Hallaey's, Schröder's and three-point methods converge slowly. If we put the initial starting point in this zone, the less number of iterations is required to reach the final solution {Brkić 2014}.

## 2.2.1 Fixed initial starting point for the Newton-Raphson method

The "center of gravity" for the "slow area" in which the Newton-Raphson method requires increased number of iterations is shown in Figure 1. The "center of gravity" has coordinates: log(Re)=4.4322→Re≈27000 and −log(ε/D)=5.7311→ε/D≈1.85·$10^{-6}$ for which the flow friction factor λ and the corresponding transmission factor $x = \frac{1}{\sqrt{\lambda}}$ can be calculated using any of the available methods. In that way calculated x, became the starting point $x_0$ for all combinations of the Reynolds number Re and the relative roughness ε/D in the domain of applicability of the Colebrook equation. With this new starting point $x_0$ the maximal required number of iterations is 4 (Figure 3), while before in the worst case was 6 (Figure 1) when the starting point $x_0$ was obtained through the "traditional formula" for this purpose; Eq. (2), all valid for the case when the flow friction factor λ is calculated with accuracy of $\lambda_{i+1}$-$\lambda_i$≤$10^{-8}$ using Colebrook equation solved in Newton's procedure when calculation goes through the transmission factor x.

The physical interpretation of this "slow area" is in the fact that this area corresponds to the initial zone of turbulent flow through smooth pipes while Eq. (2) is accurate only for the fully developed turbulent flow through rough pipes. So, Eq. (2) can already obtain accurate solution in the case of fully developed turbulent flow through rough pipes even without the iterative process, where Eq. (2) introduces the relative error of almost 80% in the case of initial phases of turbulent flow through smooth pipes.

With the initial starting point fixed at the "centre of gravity" of the "slow area", in the worst cases, maximum 4 iterations as shown in Figure 3 are enough for the required accuracy of $10^{-8}$ (before with the "traditional" version of initial value provided using Eq. (2) was 6 as indicated in Figure 1). The new fixed starting point is set as $\lambda_0$=0.024069128765100981, i.e. $x_0$=6.44569593948452. It corresponds to log(Re)=4.4322→Re≈27000 and −log(ε/D)=5.7311→ε/D≈1.85·$10^{-6}$.

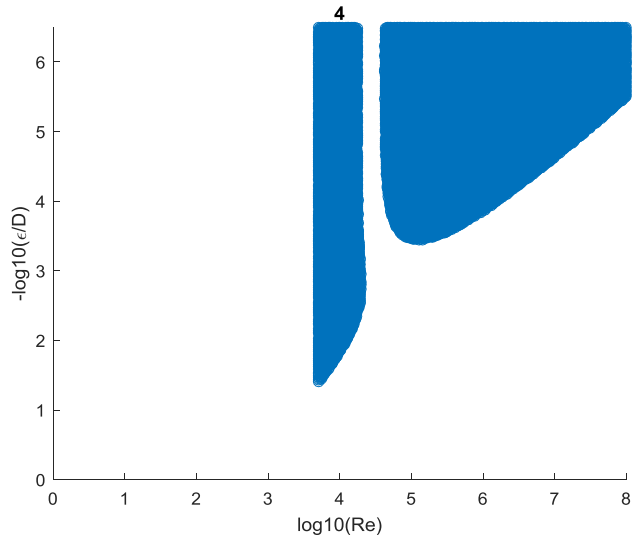

Figure 3. Decreased maximal number of required iterations from 6 to 4 to reach accuracy of $10^{-8}$ for solving Colebrook's equation in Newton's procedure when calculation goes through the transmission factor x where the initial starting point is with the fixed value: $x_0$=6.44569593948452

The new starting point $x_0=6.44569593948452$ is very robust and it seems to be an optimal starting point for all combinations when calculation goes through the transmission factor $x = \frac{1}{\sqrt{\lambda}}$ as explained in Section 3.2.

**2.2.2 Fixed initial starting point for the Halley and Schröder method**

Starting point for calculation through the Halley and Schröder method using Eq. (2) requires in the worst cases up to 4 iterations to reach the required accuracy (Figure 4). Compared with the Newton-Raphson method it is improvement of two iterations; up to 6 iterations required in Figure 1 and up to 4 iterations in Figure 4. The "worst-case" area for Halley and Schröder's method that requires 4 iterations using staring point Eq. (2) has coordinates: {$\log_{10}(Re)=5.3108 \rightarrow Re \approx 204550$; $-\log_{10}(\varepsilon/D)=4.9431 \rightarrow \varepsilon/D \approx 1.14 \cdot 10^{-5}$}$\rightarrow \lambda_0=0.015663210285978339$; i.e. $x_0=7.990256504$. This value is the new optimal initial starting point in the case of the Halley and Schröder method.

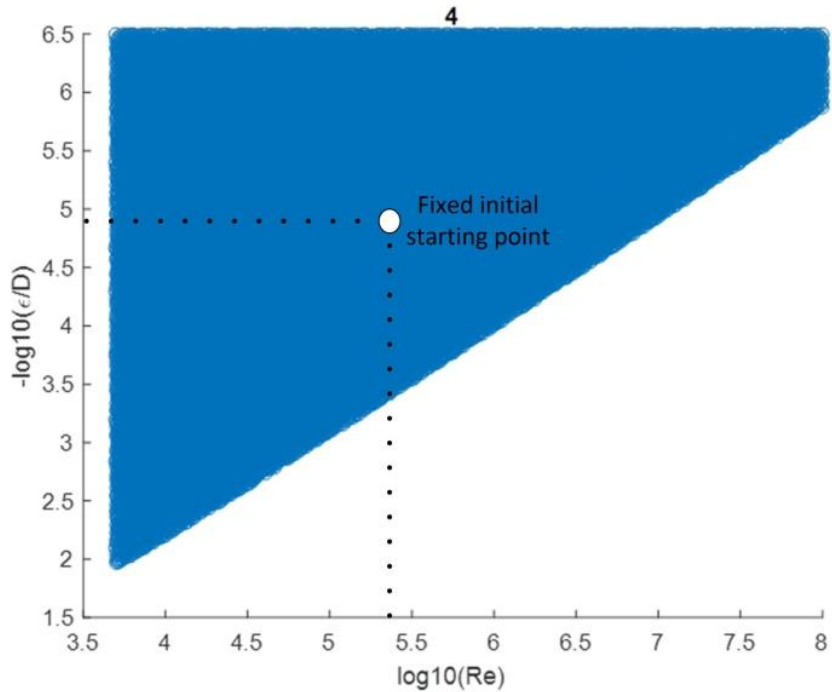

Figure 4. "Slow area" which requires 4 iterations to reach accuracy of $10^{-8}$ regarding the "traditional" option for the starting point calculated through Eq. (2), for solving Colebrook's equation in Halley's and Schröder's procedure when calculation goes through the transmission factor x

With the new initial starting point $x_0=7.990256504$, three iterations are required at maximum to reach the required accuracy in case of the Halley and Schröder method (Figure 5) as described in Section 3.2.

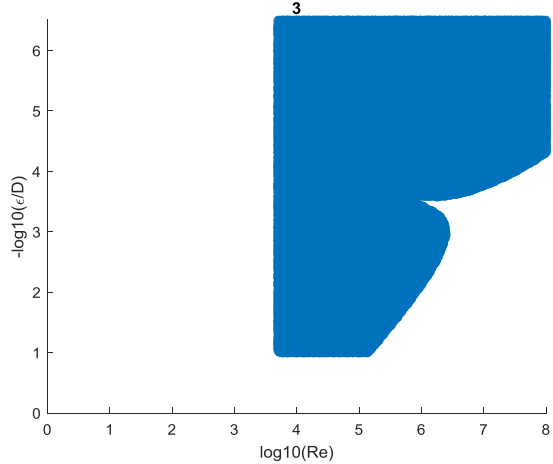

Figure 5. Decreased maximal number of required iterations from 4 to 3 to reach accuracy of $10^{-8}$ for solving Colebrook's equation in Halley's and Schröder's procedure when calculation goes through the transmission factor x where the initial starting point is with the fixed value: $x_0$=7.990256504

### 2.2.3 Fixed initial starting point for the three-point iterative methods

Optimal normalized parameters for the fixed initial starting point for the three-point iterative methods explained in Section 3.4. of this paper {Džunić et al. 2011, Petković et al. 2014, Sharma and Arora 2016} are: {$\log_{10}$(Re)=4.90060379974617→Re=79543.33576; –log(ε/D)=5.33355157079189→ε/D=4.63926·$10^{-6}$}→$\lambda_0$=0.018904186734624→$x_0$=7.273124147. The Džunić–Petković method is shown in Section 3.4. Additional recommendations about initial starting point regarding the three-point iterative methods can be found in {Yun 2008}.

### 2.3 Starting point for the Lambert W-expressed Colebrook equation

The friction factor λ in the Colebrook equation can be expressed in explicit way through the Lambert W-function { Keady 1998, Sonnad and Goudar 2006, Sonnad and Goudar 2007, Clamond 2009, Brkić 2011a, Brkić 2012b, Brkić 2017b}. The Lambert W-function further can be evaluated using some types of Householders iterative methods as shown in Section 3.5 of this paper.

The Colebrook equation in a closed form through the Lambert W-function can be expressed in two ways; Eqs. (3) and (4). The first expression is, Eq. (3) {Goudar and Sonnad 2003, Brkić 2011a, Brkić 2012b, Brkić 2017b}:

$$\frac{1}{\sqrt{\lambda}} = -2 \cdot \log_{10}\left(\frac{2 \cdot 2.51 \cdot W(y)}{Re \cdot \ln(10)} + \frac{\varepsilon}{3.7 \cdot D}\right) = -2 \cdot \log_{10}\left(10^{\frac{-W(y)}{\ln(10)}} + \frac{\varepsilon}{3.7 \cdot D}\right) \tag{3}$$

where $y = \frac{Re \cdot \ln(10)}{2 \cdot 2.51} \approx \frac{Re}{2.18}$

The argument of the Lambert W-function in this case depends only on the Reynolds number; $\lambda_0$=Re/2.18. Knowing that the practical range of the Reynolds number goes from 4000 to $10^8$, the argument of the Lambert W-function is goes from about 1835 to 45871560, where W(1835)=5.763291081 and W(45871560)=14.93748223. The Halley procedure is fast and any initial starting point can be chosen between 5 and 15, but the Newton-Raphson method is very slow and we found that for the best results the initial starting point 15 has to be chosen. Note that for W(45871560)=14.93748223, the Newton-Raphson procedure does not work in Excel for the values of initial starting point lower than 8.814.

Due to transformations of coefficients, Eq. (3) can introduce the relative error up to 2% and should be considered as explicit approximation to the Colebrook equation rather to its equivalent {Brkić 2011d}.

The second expression is; Eq. (4): {Keady 1998, Sonnad and Goudar 2004}:

$$\lambda = \left(\frac{2}{\ln(10)} \cdot W(e^\alpha) - \frac{Re \cdot \frac{\varepsilon}{D}}{3.7 \cdot 2.51}\right)^{-2} \quad (4)$$

Where $\alpha = \left(\frac{Re \cdot \frac{\varepsilon}{D} \cdot \ln(10)}{2 \cdot 2.51 \cdot 3.7} - \ln\frac{2 \cdot 3.7}{Re \cdot \ln(10)}\right)$

Argument of the Lambert W-function in this case is $e^\alpha$ which for certain combinations of Re and ε/D from the practical domain of the Colebrook equation is too big to be calculated in registers of computers {Sonad and Goudar 2004, Brkić 2012c, Vatankhah 2018}. This can be overwhelmed with the Wright Omega function; $\omega(\beta) = W(e^\alpha)$ {Wright 1959a, Wright 1959b, Corless and Jeffrey 2002, Rollmann and Spindler 2015, Biberg 2017}.

The argument of the Lambert W-function in this case, exp($\alpha_0$) depends on both Re and ε/D, but as explained due to exponential form the calculation is not always possible and because of that limited possibility of use the appropriate starting point in this case is not evaluated {Goldberg 1991, Sonnad and Goudar 2004, Kornerup and Matula 2010, Brkić 2012c}.

### 3. Iterative Methods Adopted for the Colebrook Equation

Householder's method {Householder 1970} is a numerical algorithm for solving the nonlinear equation such as Colebrook's. During the Householder's procedure, in successive calculation, i.e. in iterative cycles the original assumed value of the unknown quantity (initial starting point {Kornerup and Muller 2006}) needs to be brought as much as possible close to the real value of the quantity using the least possible number of iterations. The same situation is with the three-point methods {Sharma and Arora 2016}.

The following types of the method are used in this paper: 1[st] order Householder's method (Newton-Raphson {Cajori 1911, Ypma 1995}), 2[nd] order (Halley {Gander 1985, Gutierrez and Hernández 2001} and Schröder {Schröder 1870}), and 3[rd] order, as well three-point methods {Sharma and Arora 2016}. All these methods require calculation of the derivatives which is usually underlined as the most important shortcoming of Householder's and three-point methods compared with the simple fixed point procedure in respect of the Colebrook equation. The Newton-Raphson and three-point method require only the first derivative, the Halley and Schröder the first and the second derivative, while the 3$^{rd}$ requires the first, the second and the third derivative. In addition to the first derivate in analytical form, all required derivatives of the Colebrook function we present also in a simple and computationally inexpensive symbolic form. The derivatives in symbolic form were generated in MATLAB. In addition, the Secant method which does not require derivatives is shown as a variant of the Newton-Raphson method {Varona 2002}.

All shown approaches with Householder's methods in our case usually require only 2 to 4 iterations to reach final accurate solution {Yamamoto 2001}. This number can be slightly higher depending on the chosen method where Secant method requires by default 1-2 iterations more. Also some simple transformations of the Colebrook equation, such as introduction of the transmission factor in form of the shift $x = \frac{1}{\sqrt{\lambda}}$; can reduce number of required iterations. Knowing that the right form of equation is essential for all types of Householder's methods, here are examined two at first look very similar options: 1) direct calculation of λ in Section 3.1, and 2) indirect calculation of λ through transmission factor $x = \frac{1}{\sqrt{\lambda}}$ in Section 3.2.

Finally, the Colebrook equation can be rewritten in explicit form through the Lambert W-function {Keady 1998, Sonnad and Goudar 2004, Brkić 2012c, Brkić 2012d} and the Lambert W-function is solved in Section 3.5 using the Newton-Raphson and the Halley procedure.

### 3.1. Direct calculation of λ with derivative calculated in analytical way

The proposed technique requires the Colebrook equation in the form f(λ, Re, ε/D)=0; Eq. (5) where λ is treated as variable, the first derivative f'(λ, Re, ε/D); Eq. (6) of the Colebrook equation with respect of λ and the initial value of the friction factor $\lambda_0$ as starting point. Most probably, the function will have residue f/f'≠0 which needs to be minimized through the iterative process.

Here are the required steps for the Newton-Raphson procedure:

-The Colebrook equation in the form f(λ, Re, ε/D)=0; (5):

$$f(\lambda) = \underbrace{\frac{1}{\sqrt{|\lambda|}} + 2 \cdot \log_{10}\left(\frac{2.51}{Re \cdot \sqrt{|\lambda|}} + \frac{\varepsilon}{3.7 \cdot D}\right) = 0}_{Colebrook:\ f(\lambda)=0} \qquad (5)$$

-The first derivative f' with respect of λ in exact analytical way (6):

$$f'(\lambda) = \frac{d}{d\lambda}f(\lambda) = \underbrace{-\frac{1}{2} \cdot \left(\frac{1}{\sqrt{|\lambda|}}\right)^3 \cdot \left(1 + \frac{2 \cdot 2.51}{\ln(10) \cdot Re \cdot \left(\frac{2.51}{Re \cdot \sqrt{|\lambda|}} + \frac{\varepsilon}{3.7 \cdot D}\right)}\right)}_{1st\ derivative\ f\prime(\lambda)-Analytical} \qquad (6)$$

-Initial value $\lambda_0$ selected as explained in Section 2 of this paper in order to calculate the residue f/f' and start the iterative procedure; Eq. (7):

$$\lambda_1 = \underbrace{\lambda_0 - \frac{f(\lambda_0)}{f'(\lambda_0)}}_{Newton-Raphson} = \lambda_0 - \frac{\frac{1}{\sqrt{|\lambda_0|}}+2\cdot\log_{10}\left(\frac{2.51}{Re\cdot\sqrt{|\lambda_0|}}+\frac{\varepsilon}{3.7\cdot D}\right)}{-\frac{1}{2}\left(\frac{1}{\sqrt{|\lambda_0|}}\right)^3\cdot\left(1+\frac{2.18}{Re\cdot\left(\frac{2.51}{Re\cdot\sqrt{|\lambda_0|}}+\frac{\varepsilon}{3.7\cdot D}\right)}\right)} \qquad (7)$$

The procedure $\lambda_{i+1}=\lambda_i-f(\lambda_i)/f'(\lambda_i)$ needs to be followed until the residue $f(\lambda_i)/f'(\lambda_i)\approx 0$.

The explained Newton-Raphson procedure is shown in Tables 1 and 2 for two numerical examples: 1) {Re=5·10$^6$, ε/D=2.5·10$^{-5}$}→ λ=0.010279663295529, and 2) {Re=3·10$^4$, ε/D=9·10$^{-3}$}→λ=0.038630738574792. As explained in Section 2 of this paper, the initial starting point $\lambda_0$ in Table 1 depends on input parameters, while in Table 2 it is with the fixed value.

Table 1. Newton-Raphson procedure; Option 1: Starting point $\lambda_0$ depends on input parameters – Eq. (2), calculation of λ – Eq. (7), analytical derivative f'(λ) – Eq. (6)

| Re=5·10$^6$, ε/D=2.5·10$^{-5}$ | f(λ); Eq (5) | f'(λ); Eq. (6) | $\lambda_0$=0.009352225155363 |
|---|---|---|---|
| Iteration 1 | 0.495092014 | -573.0134258 | 0.010216239839661 |
| Iteration 2 | 0.031705666 | -502.2190127 | 0.010279370993451 |
| Iteration 3 | 0.000145453 | -497.622807 | 0.010279663289327 |
| Iteration 4 | 0.000000003 | -497.6016902 | **λ=0.010279663295529** |
| Control step | 0.000000000 | -497.6016898 | 0.010279663295529 |
| Re=3·10$^4$, ε/D=9·10$^{-3}$ | f(λ); Eq (5) | f'(λ); Eq. (6) | $\lambda_0$=0.036588313752304 |
| Iteration 1 | 0.143632267 | -73.25157738 | 0.038549121591193 |
| Iteration 2 | 0.005520057 | -67.74092562 | 0.038630609361351 |
| Iteration 3 | 0.000008725 | -67.52696208 | 0.038630738574469 |
| Iteration 4 | 0.000000000 | -67.5266237 | **λ=0.038630738574792** |
| Control step | 0.000000000 | -67.5266237 | 0.038630738574792 |

Table 2. Newton-Raphson procedure; Option 2: Fixed initial starting point $\lambda_0$=0.024069128765100981 from Section 2.2.1, calculation of $\lambda$ – Eq. (7), analytical derivative f'($\lambda$) – Eq. (6)

| Re=5·10$^6$, ε/D=2.5·10$^{-5}$ | f($\lambda$); Eq (5) | f'($\lambda$); Eq. (6) | $\lambda_0$=0.024069128765101 |
|---|---|---|---|
| Iteration 1  | -3.554956084 | -139.7424853 | -0.001370207567104 |
| Iteration 2  | 17.630891548 | -10069.59089 | 0.000380696888310 |
| Iteration 3  | 42.275315189 | -68216.8306  | 0.001000416608714 |
| Iteration 4  | 22.325487096 | -16105.99979 | 0.002386576262278 |
| Iteration 5  | 10.932300910 | -4398.30144  | 0.004872149626988 |
| Iteration 6  | 4.615550920  | -1516.202309 | 0.007916302041016 |
| Iteration 7  | 1.426053458  | -734.846953  | 0.009856914916156 |
| Iteration 8  | 0.217044469  | -529.7853757 | 0.010266598684182 |
| Iteration 9  | 0.006507144  | -498.5470019 | 0.010279650902858 |
| Iteration 10 | 0.000006167  | -497.602585  | 0.010279663295518 |
| Iteration 11 | 0.000000000  | -497.6016898 | **$\lambda$=0.010279663295529** |
| Control step | 0.000000000  | -497.6016898 | 0.010279663295529 |
| Re=3·10$^4$, ε/D=9·10$^{-3}$ | f($\lambda$); Eq (5) | f'($\lambda$); Eq. (6) | $\lambda_0$=0.024069128765101 |
| Iteration 1  | 1.391712394  | -137.1740994 | 0.034214720386916 |
| Iteration 2  | 0.326434508  | -80.9945153  | 0.038245048943635 |
| Iteration 3  | 0.026240732  | -68.54940037 | 0.038627849256271 |
| Iteration 4  | 0.000195117  | -67.53419088 | 0.038630738412914 |
| Iteration 5  | 0.000000011  | -67.52662412 | **$\lambda$=0.038630738574792** |
| Control step | 0.000000000  | -67.5266237  | 0.038630738574792 |

Here shown direct calculation of the unknown flow friction factor $\lambda$ is sensitive on the chosen initial starting point $\lambda_0$ {Kornerup and Muller 2006}. The fixed initial point $\lambda_0$ chosen as in Section 2.2 for in some cases requires increased number of iterations to reach the final solution although the procedure still maintains very good convergent properties {Yamamoto 2001, Varona 2002}. To reduce number of required iterations, use of some of the explicit approximation to the Colebrook equation are advised in order to bring the initial starting point $\lambda_0$ as close as possible near the final calculated value. Therefore, the approach with the fixed starting point as explained in Section 2.2 of this paper for all domain of the Colebrook equation, in this case cannot be advised in comparison to the approach with the starting point obtained using approximations as explained in Section 2.1.

Comparing the same approach but with the different starting points (Table 1 and 2), we can conclude that one single calculated negative value for flow friction factor $\lambda$ can increase number of required iterations significantly. These negative values can occur if the initial starting point is chosen too far away from the final calculated solution. This problem can be overwhelmed with the Colebrook function slightly rearranged as in Section 3.2.

## 3.2. Indirect calculation of λ through the transmission factor $x = \frac{1}{\sqrt{\lambda}}$

The appropriate form of the function is essential to reduce number of required iteration to reach the final solution. In order to accelerate the procedure an appropriate shift $x = \frac{1}{\sqrt{\lambda}}$ is used to provide some kind of linearization of the problem.

The Newton-Raphson procedure with these changes has similar steps as already shown:

-Shift in form of the transmission factor $x = \frac{1}{\sqrt{\lambda}}$ should be introduced in order to transform the Colebrook equation in form f(x, Re, ε/D)=0; Eq. (8):

$$f(x) = \underbrace{x + 2 \cdot \log_{10}\left(\frac{2.51 \cdot x}{Re} + \frac{\varepsilon}{3.7 \cdot D}\right) = 0}_{Colebrook:\ f(x)=0} \tag{8}$$

The first derivative of Eq. (8) in respect to the transmission factor x can be calculated analytically, but also in symbolic form (where both approaches give identical results); Sections 3.2.1 and 3.2.2.

## 3.2.1 Indirect calculation of λ through the transmission factor $x = \frac{1}{\sqrt{\lambda}}$ with the derivative calculated analytically

-The first derivative f' with respect of x can be obtained analytically; Eq. (9), (also Eq. (11) gives the same results):

$$f'(x) = \frac{d}{dx}f(x) = \underbrace{1 + 2 \cdot \frac{\frac{2.51}{Re \cdot \ln(10)}}{\frac{\varepsilon}{3.7 \cdot D} + \frac{2.51}{Re} \cdot x}}_{1st\ derivative\ f'(x) - Analytical} \tag{9}$$

-Initial value of the flow friction factor $\lambda_0$ should be chosen and the residue f/f' calculated in order to start the Newton-Raphson procedure; Eq. (10):

$$x_1 = \underbrace{x_0 - \frac{f(x_0)}{f'(x_0)}}_{Newton-Raphson} = x_0 - \frac{x_0 + 2 \cdot \log_{10}\left(\frac{2.51 \cdot x_0}{Re} + \frac{\varepsilon}{3.7 \cdot D}\right)}{1 + 2 \cdot \frac{\frac{2.51}{Re \cdot \ln(10)}}{\frac{\varepsilon}{3.7 \cdot D} + \frac{2.51}{Re} \cdot x_0}} \tag{10}$$

The procedure $x_{i+1} = x_i - f(x_i)/f'(x_i)$ should be followed until the residue $f(x_i)/f'(x_i) \approx 0$. Then the final solution is $\lambda_n = x_n^{-2}$, where n=i+1 is the final iteration.

Table 3. Newton-Raphson procedure; Option 3: fixed initial starting point
$x_0$=6.445695939→$\lambda_0$=0.024069128765101 from Section 2.2.1, indirect calculation of λ through the
transmission factor x – Eq. (10), analytical derivative f'(x) – Eq. (9)

| Re=5·10$^6$, ε/D=2.5·10$^{-5}$ | f(x); Eq. (8) | f'(x); Eq. (9) | $x_0$=6.445695939 | $\lambda_0$=0.024069128768719 |
|---|---|---|---|---|
| Iteration 1 | -3.554956085 | 1.043635910 | 9.852014225862620 | 0.010302673560706 |
| Iteration 2 | -0.011430857 | 1.037259804 | 9.863034470914730 | 0.010279663490514 |
| Iteration 3 | -0.000000097 | 1.037242198 | **x=9.863034564455800** | **λ=0.010279663295529** |
| Control step | 0.000000000 | 1.037242198 | 9.863034564455800 | 0.010279663295529 |
| Re=3·10$^4$, ε/D=9·10$^{-3}$ | f(x); Eq. (8) | f'(x); Eq. (9) | $x_0$=6.445695939 | $\lambda_0$=0.024069128768719 |
| Iteration 1 | 1.391712393 | 1.024454486 | 5.087204750239650 | 0.038640395682209 |
| Iteration 2 | -0.000651990 | 1.025427001 | 5.087840572945700 | 0.038630738577020 |
| Iteration 3 | 0.000000000 | 1.025426528 | **x=5.087840573092420** | **λ=0.038630738574792** |
| Control step | 0.000000000 | 1.046830475 | 5.087840573092420 | 0.038630738574792 |

Approach with the indirect calculation of λ through the transmission factor x is much more stable compared with the direct calculation of λ as can be seen from Tables 2 and 3 comparing the number of required iterations to reach the same accuracy (11 iterations for the direct approach compared with only 3 iterations in the indirect approach using fixed starting point $x_0$=6.445695939 for Re=5·10$^6$, ε/D=2.5·10$^{-5}$).

### 3.2.2 Indirect calculation of λ through the transmission factor $x = \frac{1}{\sqrt{\lambda}}$ with the symbolic derivative

-The exact analytical expression of the first derivative f' with respect of x can be obtained in MATLAB; Eq. (11), results are the same as using Eq. (9):

$$f'(x) = \frac{d}{dx}f(x) = \underbrace{\frac{5.02}{\text{Re}\cdot\ln(10)\cdot\left(\frac{10}{37}\cdot\frac{\varepsilon}{D}+\frac{251\cdot x}{100\cdot \text{Re}}\right)} + 1 = \frac{9287\cdot\ln(10)\cdot x + 1000\cdot\ln(10)\cdot\frac{\varepsilon}{D}\cdot Re + 18574}{\ln(10)\cdot\left(9287\cdot x + 1000\cdot\frac{\varepsilon}{D}\cdot Re\right)}}_{\text{1st derivative } f'(x) - \text{MATLAB}} \quad (11)$$

-Initial value of the flow friction factor $\lambda_0$ should be chosen and the residue f/f' calculated in order to start the Newton-Raphson procedure; Eq. (12):

$$x_1 = \underbrace{x_0 - \frac{f(x_0)}{f'(x_0)}}_{Newton-Raphson} = x_0 - \frac{x_0 + 2\cdot\log_{10}\left(\frac{2.51\cdot x_0}{Re} + \frac{\varepsilon}{3.7\cdot D}\right)}{\frac{21384.11\cdot x_0 + 2302.58\cdot\frac{\varepsilon}{D}\cdot Re + 18574}{21384.11\cdot x_0 + 2302.58\cdot\frac{\varepsilon}{D}\cdot Re}} \quad (12)$$

The procedure $x_{i+1}$=$x_i$-f($x_i$)/f'($x_i$) should be followed until the residue f($x_i$)/f'($x_i$)≈0. Then the final solution is $\lambda_n = x_n^{-2}$, where n=i+1 is the final iteration.

The iterative procedure can be accelerated using Halley's formula instead of the Newton-Raphson; Eq. (13):

$$x_1 = \underbrace{x_0 - \frac{\frac{f(x_0)}{f'(x_0)}}{1 - \frac{f(x_0) \cdot f''(x_0)}{2 \cdot (f'(x_0))^2}}}_{Halley} = \underbrace{x_0 - \frac{2 \cdot f(x_0) \cdot f'(x_0)}{2 \cdot (f'(x_0))^2 - f(x_0) \cdot f''(x_0)}}_{Halley} \tag{13}$$

In general $x_1 = x_{i+1}$ and $x_0 = x_i$; i=0 to n, where n+1 is the final iteration in which $x_n \approx x_{n+1}$.

The second derivative f''(x) in respect of x is required; Eq. (14):

$$f''(x) = \frac{d}{dx}f'(x) = \underbrace{\frac{-12.6}{Re^2 \cdot \ln(10) \cdot \left(\frac{10}{37} \cdot \frac{\varepsilon}{D} + \frac{251 \cdot x}{100 \cdot Re}\right)^2} = \frac{-172496738}{\ln(10) \cdot \left(9287 \cdot x + 1000 \cdot \frac{\varepsilon}{D} \cdot Re\right)^2}}_{\text{2nd derivative } f''(x) - MATLAB} \tag{14}$$

The Newton-Raphson method belongs to the 1$^{st}$ order, and the Halley to the 2$^{nd}$ order of Householder's method while the 3$^{rd}$ order can be expressed using Eq. (15):

$$x_1 = \underbrace{x_0 - \frac{6 \cdot f(x_0) \cdot (f'(x_0))^2 - 3 \cdot (f(x_0))^2 \cdot f''(x_0)}{6 \cdot (f'(x_0))^3 - 6 \cdot f(x_0) \cdot f'(x_0) \cdot f''(x_0) + (f(x_0))^2 \cdot f'''(x_0)}}_{\text{3rd order Householder}} \tag{15}$$

Again, $x_1 = x_{i+1}$ and $x_0 = x_i$; i=0 to n, where n+1 is final iteration in which $x_n \approx x_{n+1}$.

The required 3$^{rd}$ derivative f'''(x) can be expressed using Eq. (16):

$$f'''(x) = \frac{d}{dx}f''(x) = \underbrace{\frac{63.253}{Re^3 \cdot \ln(10) \cdot \left(\frac{10}{37} \cdot \frac{\varepsilon}{D} + \frac{251 \cdot x}{100 \cdot Re}\right)^3} = \frac{3203954411612}{\ln(10) \cdot \left(9287 \cdot x + 1000 \cdot \frac{\varepsilon}{D} \cdot Re\right)^3}}_{\text{3rd derivative } f'''(x) - MATLAB} \tag{16}$$

Also here one has to be underlined that the Halley method {Brown 1977} is not the unique Householder's method of the 2$^{nd}$ order {Householder 1970}. For example Schröder's method {Schröder 1870} belongs also to the group; Eq. (17):

$$x_1 = \underbrace{x_0 - \frac{f(x_0)}{f'(x_0)} - \frac{f''(x_0) \cdot (f(x_0))^2}{2 \cdot (f'(x_0))^3}}_{Schröder} \tag{17}$$

Further $x_1 = x_{i+1}$ and $x_0 = x_i$; i=0 to n, where n+1 is final iteration in which $x_n \approx x_{n+1}$.

Using the presented Householder procedures; 1$^{st}$ order: the Newton-Raphson, 2$^{nd}$: Halley, and 3$^{rd}$, the unknown flow friction factor λ should be calculated for the two given pairs of the Reynolds number Re and the relative roughness ε/D: 1) {Re=5·10$^6$, ε/D=2.5·10$^{-5}$}→ λ=0.010279663295529, and 2) {Re=3·10$^4$, ε/D=9·10$^{-3}$}→λ=0.038630738574792. The calculation presented in Tables 4-7 is through the transmission factor x, using the symbolic derivative f'(x), but with different initial starting point λ$_0$.

Table 4. Newton-Raphson procedure; Option 4: starting point $x_0$ depends on input parameters – Eq. (2), indirect calculation of λ through the transmission factor x – Eq. (12), the symbolic derivative f'(x) – Eq. (11)

| Re=5·10$^6$, ε/D=2.5·10$^{-5}$ | f(x); Eq. (8) | f'(x); Eq. (11) | $x_0$=10.34052343 | $λ_0$=0.009352225155363 |
|---|---|---|---|---|
| Iteration 1 | 0.495092014 | 1.036495031 | 9.862863625818000 | 0.010280019623455 |
| Iteration 2 | -0.000177305 | 1.037242471 | 9.863034564433310 | 0.010279663295576 |
| Iteration 3 | 0.000000000 | 1.037242198 | **x=9.863034564455800** | **λ=0.010279663295529** |
| Control step | 0.000000000 | 1.037242198 | 9.863034564455800 | 0.010279663295529 |
| Re=3·10$^4$, ε/D=9·10$^{-3}$ | f(x); Eq. (8) | f'(x); Eq. (11) | $x_0$=5.227918429 | $λ_0$=0.036588313752304 |
| Iteration 1 | 0.143632267 | 1.025322691 | 5.087833489750430 | 0.038630846139210 |
| Iteration 2 | -0.000007263 | 1.025426533 | 5.087840573092400 | 0.038630738574793 |
| Iteration 3 | 0.000000000 | 1.025426528 | **x=5.087840573092420** | **λ=0.038630738574792** |
| Control step | 0.000000000 | 1.025426528 | 5.087840573092420 | 0.038630738574792 |

Table 5. Halley procedure; Option 5: fixed initial starting point $x_0$=7.990256504→$λ_0$=0.015663210285978339 from Section 2.2.2, indirect calculation of λ through the transmission factor x – Eq. (13), the symbolic derivatives f'(x) and f''(x) – Eqs. (11) and (14)

| Re=5·10$^6$, ε/D=2.5·10$^{-5}$ | f(x); Eq. (8) | f'(x); Eq. (11) | f''(x); Eq. (14) | $x_0$=7.990256504 | $λ_0$=0.015663210285978 |
|---|---|---|---|---|---|
| Iteration 1 | -1.945484250 | 1.040493788 | -0.001887828 | 9.863203600915390 | 0.010279310950983 |
| Iteration 2 | 0.000175332 | 1.037241928 | -0.001596798 | **x=9.863034564455800** | **λ=0.010279663295529** |
| Control step | 0.000000000 | 1.037242198 | -0.001596821 | 9.863034564455800 | 0.010279663295529 |
| Re=3·10$^4$, ε/D=9·10$^{-3}$ | f(x); Eq. (8) | f'(x); Eq. (11) | f''(x); Eq. (14) | $x_0$=7.990256504 | $λ_0$=0.015663210285978 |
| Iteration 1 | 2.973246188 | 1.023435376 | -0.000632309 | 5.087698791122220 | 0.038632891696967 |
| Iteration 2 | -0.000145387 | 1.025426633 | -0.000744326 | **x=5.087840573092420** | **λ=0.038630738574792** |
| Control step | 0.000000000 | 1.025426528 | -0.000744320 | 5.087840573092420 | 0.038630738574792 |

Table 6. 3$^{rd}$ order Householder's procedure; Option 6: starting point $x_0$ depends on input parameters – Eq. (2), indirect calculation of λ through the transmission factor x – Eq. (15), the symbolic derivatives f'(x), f''(x) and f'''(x) – Eqs. (11), (14) and (16)

| Re=5·10$^6$, ε/D=2.5·10$^{-5}$ | f(x); Eq. (8) | f'(x); Eq. (11) | f''(x); Eq. (14) | f'''(x); Eq. (16) | $x_0$=10.34052343 | $λ_0$=0.009352225155363 |
|---|---|---|---|---|---|---|
| Iteration 1 | 0.495092014 | 1.036495031 | -0.001533392 | 0.000128855 | 9.863034531578420 | 0.010279663364062 |
| Iteration 2 | -0.000000034 | 1.037242198 | -0.001596821 | 0.000136933 | **x=9.863034564455800** | **λ=0.010279663295529** |
| Control step | 0.000000000 | 1.037242198 | -0.001596821 | 0.000136933 | 9.863034564455800 | 0.010279663295529 |
| Re=3·10$^4$, ε/D=9·10$^{-3}$ | f(x); Eq. (8) | f'(x); Eq. (11) | f''(x); Eq. (14) | f'''(x); Eq. (16) | $x_0$=10.34052343 | $λ_0$=0.009352225155363 |
| Iteration 1 | 0.143632267 | 1.025322691 | -0.000738253 | 0.000043046 | 5.087840573035260 | 0.038630738575660 |
| Iteration 2 | 0.000000000 | 1.025426528 | -0.000744320 | 0.000043578 | **x=5.087840573092420** | **λ=0.038630738574792** |
| Control step | 0.000000000 | 1.025426528 | -0.000744320 | 0.000043578 | 5.087840573092420 | 0.038630738574792 |

Table 7. Schröder procedure; Option 7: fixed initial starting point $x_0$=7.990256504→$λ_0$=0.015663210285978339 from Section 2.2.2, indirect calculation of λ through the transmission factor x – Eq. (17), the symbolic derivatives f'(x) and f''(x) – Eqs. (11) and (14)

| Re=5·10$^6$, ε/D=2.5·10$^{-5}$ | f(x); Eq. (8) | f'(x); Eq. (11) | f''(x); Eq. (14) | $x_0$=7.990256504 | $λ_0$=0.015663210285978 |
|---|---|---|---|---|---|
| Iteration 1 | -1.945484250 | 1.040493788 | -0.001887828 | 9.863198212166060 | 0.010279322183170 |
| Iteration 2 | 0.000169742 | 1.037241937 | -0.001596799 | **x=9.863034564455800** | **λ=0.010279663295529** |
| Control step | 0.000000000 | 1.037242198 | -0.001596821 | 9.863034564455800 | 0.010279663295529 |
| Re=3·10$^4$, ε/D=9·10$^{-3}$ | f(x); Eq. (8) | f'(x); Eq. (11) | f''(x); Eq. (14) | $x_0$=7.990256504 | $λ_0$=0.015663210285978 |
| Iteration 1 | 2.973246188 | 1.023435376 | -0.000632309 | 5.087701128882780 | 0.038632856193927 |
| Iteration 2 | -0.000142990 | 1.025426632 | -0.000744326 | **x=5.087840573092420** | **λ=0.038630738574792** |
| Control step | 0.000000000 | 1.025426528 | -0.000744320 | 5.087840573092420 | 0.038630738574792 |

### 3.3. Secant method

Secant method is similar to the Newton-Raphson, it requires two starting points $\lambda_0$ and $\lambda_{-1}$ but doesn't require calculation of the derivatives {Varona 2002}. The approach with the direct calculation of $\lambda$ with the two required starting points $\lambda_0$ and $\lambda_{-1}$ is formulated; Eq. (18):

$$\lambda_1 = \lambda_0 - \frac{f(\lambda_0)}{\frac{f(\lambda_{-1})-f(\lambda_0)}{\lambda_{-1}-\lambda_0}} \tag{18}$$

Counter i starts from i-1 and goes to n+1 in which $\lambda_n=\lambda_{n+1}$.

The approach through the transmission factor x is; Eq. (19):

$$x_1 = x_0 - \frac{f(x_0)}{\frac{f(x_{-1})-f(x_0)}{x_{-1}-x_0}} \tag{19}$$

As already described, counter i also starts from i-1 and goes to n+1 in which $x_n=x_{n+1}$.

Flow friction factor $\lambda$ is calculated in Tables 8 and 9 for two pairs of the Reynolds number and the relative roughness 1) Re=5·10$^6$, ε/D=2.5·10$^{-5}$ and 2) Re=3·10$^4$, ε/D=9·10$^{-3}$ using the Secant procedure with direct calculation of $\lambda$ and indirect through the transmission factor x.

Table 8. Secant procedure; Option 8: two initial starting points $\lambda_0$ and $\lambda_{-1}$ required: starting point $\lambda_{-1}$ is with fixed value $\lambda_{-1}$=0.024069128765101 (i.e. $x_{-1}$=6.445695939) as in Section 2.2.1, while starting point $\lambda_0$ depends on input parameters – Eq. (2), direct calculation of $\lambda$ – Eq. (18)

| Re=5·10$^6$, ε/D=2.5·10$^{-5}$ | $f(\lambda_{i-1})$; Eq. (5) | $f(\lambda_i)$; Eq. (5) | $\frac{f(\lambda_{i-1})-f(\lambda_i)}{\lambda_{i-1}-\lambda_i}$ | $\lambda_{-1}$=0.024069128765101 $\lambda_0$=0.009352225155363 |
|---|---|---|---|---|
| Iteration 1 | 0.495092014 | -3.554956084 | -275.1970255 | 0.011151270814558 |
| Iteration 2 | -0.408071981 | 0.495092014 | -502.0239429 | 0.010338417191085 |
| Iteration 3 | -0.029111936 | -0.408071981 | -466.2094551 | 0.010275973292109 |
| Iteration 4 | 0.001836644 | -0.029111936 | -495.6221591 | 0.010279679026163 |
| Iteration 5 | -0.000007828 | 0.001836644 | -497.734448 | 0.010279663299743 |
| Iteration 6 | -0.000000002 | -0.000007828 | -497.6011214 | **λ=0.010279663295529** |
| Control step | 0.000000000 | -0.000000002 | -497.6012179 | 0.010279663295529 |
| Re=3·10$^4$, ε/D=9·10$^{-3}$ | $f(\lambda_{i-1})$; Eq. (5) | $f(\lambda_i)$; Eq. (5) | $\frac{f(\lambda_{i-1})-f(\lambda_i)}{\lambda_{i-1}-\lambda_i}$ | $\lambda_{-1}$=0.024069128765101 $\lambda_0$=0.036588313752304 |
| Iteration 1 | 1.391712394 | 0.143632267 | -99.69340079 | 0.038029053721052 |
| Iteration 2 | 0.143632267 | 0.041110009 | -71.15944585 | 0.038606770549177 |
| Iteration 3 | 0.041110009 | 0.001619232 | -68.35663251 | 0.038630458556837 |
| Iteration 4 | 0.001619232 | 0.000018909 | -67.5583902 | 0.038630738444645 |
| Iteration 5 | 0.000018909 | 0.000000009 | -67.52699052 | **λ=0.038630738574792** |
| Control step | 0.000000009 | 0.000000000 | -67.52662212 | 0.038630738574792 |

Table 9. Secant procedure; Option 9: two initial starting points $x_0$ and $x_{-1}$ required: starting point $\lambda_{-1}$ is with fixed value $x_{-1}=6.445695939$ (i.e. $\lambda_{-1}=0.024069128765101$) as in Section 2.2.1; while starting point $\lambda_0$ depends on input parameters – Eq. (2), indirect calculation of λ through the transmission factor x – Eq. (19)

| Re=5·10⁶, ε/D=2.5·10⁻⁵ | $f(x_{i-1})$; Eq. (8) | $f(x_i)$; Eq. (8) | $\frac{f(x_{i-1})-f(x_i)}{x_{i-1}-x_i}$ | $x_{-1}$=6.445695939 $x_0$=10.34052343 | $\lambda_{-1}$=0.024069128765101 $\lambda_0$=0.009352225155363 |
|---|---|---|---|---|---|
| Iteration 1 | -3.554956084 | 0.495092014 | 1.039853012 | 9.864406125318800 | 0.010276804896656 |
| Iteration 2 | 0.495092014 | 0.001422639 | 1.03686501 | 9.863034066961850 | 0.010279664332547 |
| Iteration 3 | 0.001422639 | -0.000000516 | 1.037241104 | 9.863034564456330 | 0.010279663295528 |
| Iteration 4 | -0.000000516 | 0.000000000 | 1.0372422 | **x=9.863034564455800** | **λ=0.010279663295529** |
| Control step | 0.000000000 | 0.000000000 | 1.037162162 | 9.863034564455800 | 0.010279663295529 |
| Re=3·10⁴, ε/D=9·10⁻³ | $f(x_{i-1})$; Eq. (8) | $f(x_i)$; Eq. (8) | $\frac{f(x_{i-1})-f(x_i)}{x_{i-1}-x_i}$ | $x_{-1}$=6.445695939 $x_0$=5.227918429 | $\lambda_{-1}$=0.024069128765101 $\lambda_0$=0.036588313752304 |
| Iteration 1 | 1.391712394 | 0.143632267 | 1.024883541 | 5.087773465040530 | 0.038631757665255 |
| Iteration 2 | 0.143632267 | -0.000068814 | 1.025374564 | 5.087840576494990 | 0.038630738523123 |
| Iteration 3 | -0.000068814 | 0.000000003 | 1.025426553 | **x=5.087840573092420** | **λ=0.038630738574792** |
| Control step | 0.000000003 | 0.000000000 | 1.025426591 | 5.087840573092420 | 0.038630738574792 |

Calculation using the Secant procedure, also confirms that the indirect calculation of λ through the transmission factor x requires in general less number of iterations to reach the same level of accuracy. In the case from Tables 8 and 9, the required number of iterations is 6 in direct calculation and 5 in indirect for Re=5·10$^6$, ε/D=2.5·10$^{-5}$, the required number of iterations is 4 in direct calculation and 3 in indirect for Re=3·10$^4$, ε/D=9·10$^{-3}$.

### 3.4. Three-point methods

Here we will apply the Džunić–Petković three-point iterative method to the Colebrook equation {Džunić et al. (2011), Sharma and Arora (2016)}. It requires only one iteration to reach final accurate solution {Brkić and Praks 2019} and we will show all steps to calculate the friction factor λ; Eq. (20) where the numerical values are for Re=5·10$^6$, ε/D=2.5·10$^{-5}$. Initial starting point is $x_0$=7.273124147 as described in Section 2.2.3.

$$\left.\begin{array}{c} x_0 = 7.273124147 \\ f(x_0) = x_0 + 2 \cdot \log_{10}\left(\frac{2.51 \cdot x_0}{Re} + \frac{\varepsilon}{3.7 \cdot D}\right) = -2.692152546 \\ f'(x_0) = \frac{9287 \cdot \ln(10) \cdot x_0 + 1000 \cdot \ln(10) \cdot \frac{\varepsilon}{D} \cdot Re + 18574}{\ln(10) \cdot \left(9287 \cdot x_0 + 1000 \cdot \frac{\varepsilon}{D} \cdot Re\right)} = \frac{5.02}{Re \cdot \ln(10) \cdot \left(\frac{10}{37} \cdot \frac{\varepsilon}{D} + \frac{2.51 \cdot x_0}{Re}\right)} + 1 = 1.041894438 \\ y_0 = x_0 - \frac{f(x_0)}{f'(x_0)} = 7.273124147 - \frac{-2.692152546}{1.041894} = 9.857025593360860 \\ f(y_0) = y_0 + 2 \cdot \log_{10}\left(\frac{2.51 \cdot y_0}{Re} + \frac{\varepsilon}{3.7 \cdot D}\right) = -0.006232787 \\ z_0 = y_0 - \frac{f(x_0)}{f(x_0) - 2 \cdot f(y_0)} \cdot \frac{f(y_0)}{f'(x_0)} = 9.863035589 \\ f(z_0) = z_0 + 2 \cdot \log_{10}\left(\frac{2.51 \cdot z_0}{Re} + \frac{\varepsilon}{3.7 \cdot D}\right) = -0.006232787 \\ x_1 = z_0 - \frac{f(z_0)}{f'(x_0) \cdot \left[1 - 2 \cdot \frac{f(y_0)}{f(x_0)} - \left(\frac{f(y_0)}{f(x_0)}\right)^2\right] \cdot \left[1 - \frac{f(z_0)}{f(y_0)}\right] \cdot \left[1 - 2 \cdot \frac{f(z_0)}{f(x_0)}\right]} = 9.863034564 \\ x_1 = 9.863034564 \rightarrow \lambda_1 = 0.010279663295529 \end{array}\right\} \quad (20)$$

### 3.5. Expressed through the Lambert W-function

The Lambert W-function W(y) {Corless et al. 1996} is the solution of $y = z \cdot e^z$, which needs to be in the appropriate form; Eq. (21):

$$f(z) = z \cdot e^z - y = 0 \tag{21}$$

-The first derivative f'(z); Eq. (22):

$$f'(z) = e^z \cdot (z + 1) \tag{22}$$

-Choose initial value $z_0$, calculate the residue f/f' and start the procedure; Eq. (23):

$$Z_1 = Z_0 - \frac{f(z_0)}{f'(z_0)} = \underbrace{Z_0 - \frac{z_0 \cdot e^{z_0} - y}{e^{z_0} \cdot (z_0+1)}}_{Newton-Raphson} = Z_0 - \frac{z_0 \cdot e^{z_0} - \frac{Re}{2.18}}{e^{z_0} \cdot (z_0+1)} \tag{23}$$

Then follow the procedure $z_{i+1}=z_i-f(z_i)/f'(z_i)$ until the residue $f(z_i)/f'(z_i) \approx 0$, where n=i+1 in final iteration.

Halley's procedure; Eq. (24):

$$Z_1 = Z_0 - \frac{f(z_0)}{f'(z_0) - \frac{f(z_0) \cdot f''(z_0)}{2 \cdot f'(z_0)}} = \underbrace{Z_0 - \frac{z_0 \cdot e^{z_0} - y}{e^{z_0} \cdot (z_0+1) - \frac{(z_0 \cdot e^{z_0} - y) \cdot (z_0+2)}{2 \cdot (z_0+1)}}}_{Halley} \tag{24}$$

Where the second derivative is; Eq. (25):

$$f''(z) = e^z \cdot (z+2) \tag{25}$$

The Schröder expression is; Eq. (26):

$$Z_1 = Z_0 - \frac{f(z_0)}{f'(z_0)} - \frac{f''(z_0) \cdot (f(z_0))^2}{2 \cdot (f'(z_0))^3} = \underbrace{Z_0 - \frac{z \cdot e^z - y}{e^z \cdot (z+1)} - \frac{e^z \cdot (z+2) \cdot (z \cdot e^z - y)^2}{2 \cdot (e^z \cdot (z+1))^3}}_{Schröder} \tag{26}$$

Further in all cases, the Newton-Raphson, the Halley, and the Schröder; $z_1=z_{i+1}$ and $z_0=z_i$; i=0 to n, where n+1 is final iteration in which $z_n \approx z_{n+1}$.

Argument of the Lambert W-function y, in our case defined by Eq. (3), is y=Re/2.18. Therefore it does not depend on the relative roughness ε/D but only on the Reynolds number Re. In Table 10, z is calculated in the iterative procedure using the Newton-Raphson and Halley method for Re=5·10⁶ and Re=3·10⁴ where initial starting point is set as $z_0$=15 as recommended in Section 2.3 of this paper (for $z_0$<8.814, the Newton-Raphson procedure cannot start).

Table 10. Calculation of W(y) where y=Re/2.18 using the Newton-Raphson, the Halley and the Schröder iterative methods

|  | Re=5·10⁶, y=Re/2.18=2293411.45 | | | Re=3·10⁴, y=Re/2.18=13760.47 | | |
|---|---|---|---|---|---|---|
|  | Newton-Raphson | Halley | Schröder | Newton-Raphson | Halley | Schröder |
| Iteration 0 | $z_0$=15 | $z_0$=15 | $z_0$=15 | $z_0$=15 | $z_0$=15 | $z_0$=15 |
| Iteration 1 | 14.10634749 | 13.29860556 | 13.68208338 | 14.06276308 | 13.1333396 | 13.59610616 |
| Iteration 2 | 13.28604947 | 12.2757343 | 12.62556802 | 13.12986539 | 11.29171838 | 12.20340124 |
| Iteration 3 | 12.62863905 | 12.14855784 | 12.17738057 | 12.20257063 | 9.520829163 | 10.83006437 |
| Iteration 4 | 12.25343232 | **z=12.14835704** | 12.14836628 | 11.28354302 | 8.068323472 | 9.505729616 |
| Iteration 5 | 12.15407754 | 12.14835704 | **z=12.14835704** | 10.37904335 | 7.530266826 | 8.341483562 |
| Iteration 6 | 12.14837461 |  | 12.14835704 | 9.504505014 | 7.512930233 | 7.637280252 |
| Iteration 7 | **z=12.14835704** |  |  | 8.697314341 | **z=7.512929679** | 7.513654122 |
| Iteration 8 | 12.14835704 |  |  | 8.037456295 | 7.512929679 | **z=7.512929679** |
| Iteration 9 |  |  |  | 7.640105762 |  | 7.512929679 |
| Iteration 10 |  |  |  | 7.52154464 |  |  |
| Iteration 11 |  |  |  | 7.512971011 |  |  |
| Iteration 12 |  |  |  | 7.51292968 |  |  |
| Iteration 13 |  |  |  | **z=7.512929679** |  |  |
| Iteration 14 |  |  |  | 7.512929679 |  |  |

## 4. Approximations – Simplified equations for engineering practice

Using optimal fixed initial starting point for the Halley and the Schröder method as explained in Section 2.2.2, the first iteration of the procedures from Section 3.2.2, simplification using the fact that the first derivative of the Colebrook function is almost always near zero; f'≈0; and using acceleration through Eq. (28) {Zigrang and Sylvester 1982, Serghides 1984}, the following approximations; Eq. (27) can be formed. Using Eq. (27), the maximal relative error in the domain of applicability of the Colebrook equation is 8.29% (Figure 6), and using acceleration Eq. (28), i.e. single fixed point iterative method {Brkić 2017a}, the maximal relative error is 0.69% (Figure 7), 0.0617% (Figure 8), etc.

$$\frac{1}{\sqrt{\lambda_0}} \approx \underbrace{8 - \frac{2 \cdot A}{2 - A \cdot B}}_{Halley} \approx \underbrace{8 - A - \frac{A^2 \cdot B}{2}}_{Schröder} \approx \underbrace{8 - \frac{6 \cdot A - 3 \cdot A^2 \cdot B}{6 - 6 \cdot A \cdot B + A^2 \cdot C}}_{3^{rd}\ order} \tag{27}$$

Then $\lambda_0$ from Eq. (27) is used in Eq. (28):

$$\left. \begin{array}{l} \frac{1}{\sqrt{\lambda_1}} \approx \underbrace{-2 \cdot \log_{10}\left(\frac{2.51}{Re} \cdot \frac{1}{\sqrt{\lambda_0}} + \frac{\varepsilon}{3.7 \cdot D}\right)}_{1st\ Colebrook's\ acceleration} \\[2ex] \frac{1}{\sqrt{\lambda_2}} \approx \underbrace{-2 \cdot \log_{10}\left(\frac{2.51}{Re} \cdot \frac{1}{\sqrt{\lambda_1}} + \frac{\varepsilon}{3.7 \cdot D}\right)}_{2st\ Colebrook's\ acceleration} = -2 \cdot \log_{10}\left(\frac{2.51}{Re} \cdot \underbrace{\left(-2 \cdot \log_{10}\left(\underbrace{\frac{2.51}{Re} \cdot \frac{1}{\sqrt{\lambda_0}} + \frac{\varepsilon}{3.7 \cdot D}}_{1st\ Colebrook's\ acceleration}\right)\right)}_{2st\ Colebrook's\ acceleration} + \frac{\varepsilon}{3.7 \cdot D}\right) \\[2ex] \vdots \\[2ex] \frac{1}{\sqrt{\lambda_{i+1}}} \approx \underbrace{-2 \cdot \log_{10}\left(\frac{2.51}{Re} \cdot \frac{1}{\sqrt{\lambda_i}} + \frac{\varepsilon}{3.7 \cdot D}\right)}_{Colebrook-acceleration} \end{array} \right\} \tag{28}$$

Where A, B, and C are; Eqs. (29-31)

$$A \approx 8 + 2 \cdot \log_{10}\left(\frac{16}{Re} + \frac{\varepsilon}{3.7 \cdot D}\right) \tag{29}$$

$$B \approx \frac{-74914381.46}{\nabla^2} \tag{30}$$

$$C \approx \frac{1391459721232.67}{\nabla^3} \tag{31}$$

Where $\nabla$; Eq. (32)

$$\nabla \approx 74205.5 + 1000 \cdot \frac{\varepsilon}{D} \cdot Re \tag{32}$$

The shown procedure is efficient and does not require extensive computing resources since the accuracy of more than 0.69% can be reached using only two logarithmic forms, while very high accuracy of 0.0617% using only three logarithmic forms (in all cases exponents are only whole numbers) {Clamond 2009, Giustolisi et al. 2011, Vatankhah 2018}.

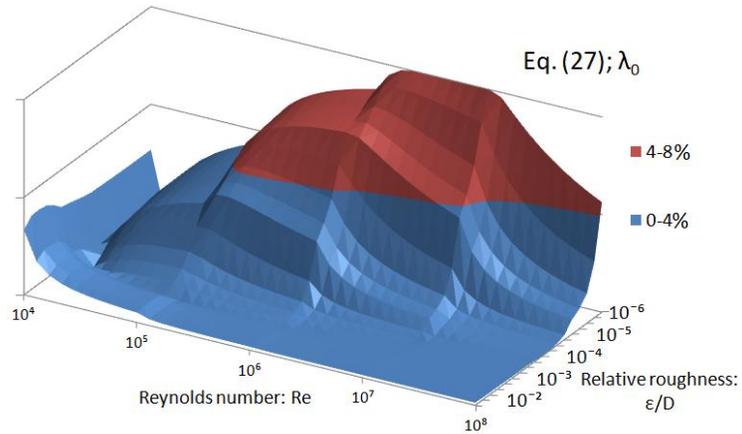

Figure 6. Distribution of the relative error over the domain of applicability of the Colebrook equation of the approximation; Eq. (27) for i=0 the maximal relative error is 8.29%

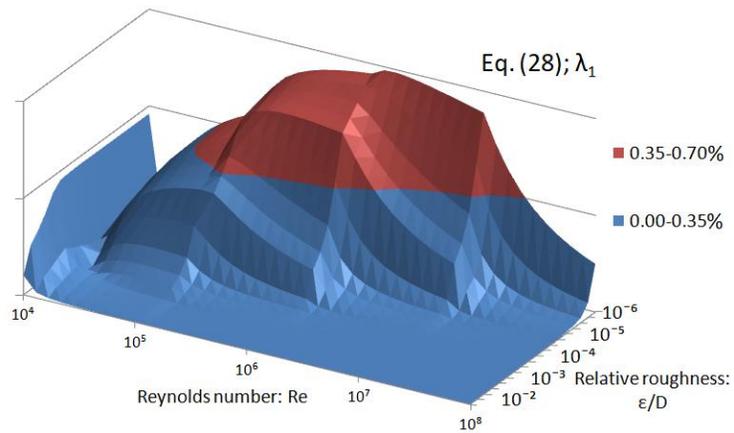

Figure 7. Distribution of the relative error over the domain of applicability of the Colebrook equation of the approximation; Eq. (28) for i=0 the maximal relative error is 0.69%

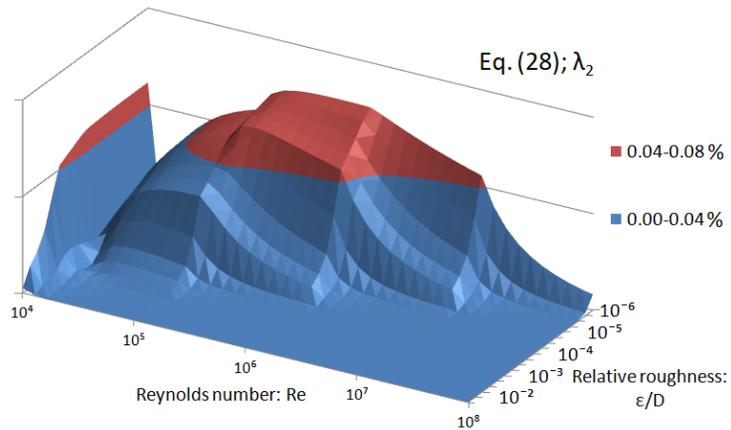

Figure 8. Distribution of the relative error over the domain of applicability of the Colebrook equation of the approximation; Eq. (28) for i=1 the maximal relative error is 0.0617%

The error analysis is in the domain of applicability of the Colebrook equation and can be further reduced using one more accelerating step as shown, genetic algorithms {Ćojbašić and Brkić 2013, Bartz-Beielstein et al. 2014, Brkić and Ćojbašić 2017}, Excel fitting tool {Vatankhah 2014} or Monte Carlo {Praks et al. 2015, Praks et al. 2017}.

## 5. Conclusion

Today, a fast but reliable approximation of pipeline hydraulic is needed also reliability modelling of water and gas networks where a large number of network simulations of random component failures and their combinations must be automatically evaluated and statistically analyzed {Brkić 2009, Spiliotis and Tsakiris 2010, Brkić 2011e, Brkić 2011f, Brkić 2012e, Brkić 2016b, Brkić 2018}. Accurate, fast and reliable estimation of flow friction factors are essential for evaluation of pressure drops and flows in large network of pipes, because, for example, compression station failure can be approximated on the transmission level by user-defined logic rules obtained from hydraulic software {Brkić and Tanasković 2008, Praks et al. 2015, Praks et al. 2017, Badami et al. 2018, Tran et al. 2018}. Iterative solutions and approximations for calculation of flow friction factor are implemented in software packages which are in common use in everyday engineering practice {Brkić 2016b}. So in this paper we analyzed selected iterative procedures in order to solve the Colebrook equation {Chun and Neta 2017, Zhanlav et al. 2017} and we found that two or three iterations of Halley and the Schröder methods are suitable for the required accuracy needed for the engineering practice, when the fixed initial starting point of Section 2.2.2 is applied. On the other hand, using three-point iterative method with the same initial conditions, the required high accuracy can be reached after only one iteration but using three internal steps {Džunić et al. 2011, Petković et al. 2014, Sharma and Arora 2016}.

Moreover, to simplified calculation for engineering use we can recommend:

1. Knowing that the Colebrook equation is used in engineering practice only in the limited domain of the Reynolds number Re between 4000 and $10^8$, and of the relative roughness of inner pipe surface ε/D up to 0.05, we evaluated number of iteration from that domain to reach sufficient accuracy and we detected zones in which additional number of iterations are required. Therefore we put the fixed initial point to start the iterative procedure in that zones in order to decrease number of required iterations.
2. Using the simplified Halley and the simplified Schröder procedure with the fixed starting point, after only one iteration we developed approximation that is with the error up to 8.29%. This is near the accurate value and therefore for the second iteration the simplified Newton-Raphson method can be used to reach accuracy of 0.69% and for extreme accuracy the third iteration need to be used to reach accuracy of 0.0617%. The simplification is in fact that the first derivative of the Colebrook function in our case is always near zero; f'≈0 (for f'→0 the Newton-Raphson method become the fixed-point method). Accuracy of 0.69% can be reached using only two logarithmic forms and of 0.0617% with only three logarithmic forms which does not require extensive computational efforts (the goal is to use the least possible number of logarithmic functions or functions with non-integer power) {Giustolisi et al 2011, Vatankhah 2018}. Moreover, the computational cost of iterations can be also reduced using Padé polynomials {Praks and Brkić 2018}.

We analyzed also the Colebrook equation expressed through the Lambert W-function and we found that the Halley and the Schröder methods can be advised in comparison to the Newton-Rapson method (where the problem with the initial starting point exists).

**Conflicts of Interest**

The authors declare that there is no conflict of interest regarding the publication of this paper. The views expressed are those of the authors and may not in any circumstances be regarded as stating an official position of the affiliated employers of the authors; European Commission and Technical University Ostrava. Both authors contributed equally to this study.

**Data availability statement**

All conclusions are based on the calculation shown in this paper. All data which is required to repeat the calculation can be obtained using equation shown in this paper. Examples are also shown in Tables to support conclusions.

Goudar, C.T. and Sonnad, J.R. (2003). Explicit friction factor correlation for turbulent flow in smooth pipes. Industrial & Engineering Chemistry Research 42(12), 2878-2880. https://dx.doi.org/10.1021/ie0300676

Gregory, G.A. and Fogarasi, M. (1985). Alternate to standard friction factor equation. Oil & Gas Journal 83(13), 120 and 125–127.

Griffiths, D.V. and Smith, I.M. (2006). Numerical methods for engineers. CRC press.

Gutierrez, J.M. and Hernández, M.A. (2001). An acceleration of Newton's method: Super-Halley method. Applied Mathematics and Computation 117(2-3), 223-239. https://doi.org/10.1016/S0096-3003(99)00175-7

Halley, E. (1694). A new, exact and easy method of finding the roots of equations generally, and that without any previous reduction. Philosophical Transactions of the Royal Society 18(210), 136-148. doi:10.1098/rstl.1694.0029

Hayes, B. (2005). Why W? American Scientist 93(2), 104–108. https://dx.doi.org/10.1511/2005.2.104

Herwig, H., Gloss, D. and Wenterodt, T. (2008). A new approach to understanding and modelling the influence of wall roughness on friction factors for pipe and channel flows. Journal of Fluid Mechanics 613, 35-53. https://doi.org/10.1017/S0022112008003534

Hosseini, M., Chizari, H., Poston, T., Salleh, M.B. and Abdullah, A.H. (2014). Efficient underwater RSS value to distance inversion using the Lambert function. Mathematical Problems in Engineering http://dx.doi.org/10.1155/2014/175275

Householder, A.S. (1970). The Numerical Treatment of a Single Nonlinear Equation. McGraw-Hill. https://lccn.loc.gov/79103908

Keady, G. (1998). Colebrook-White formula for pipe flows. Journal of Hydraulic Engineering ASCE 124(1), 96-97. https://dx.doi.org/10.1061/(ASCE)0733-9429(1998)124:1(96)

Kornerup, P. and Matula, D.W. (2010). Finite precision number systems and arithmetic. Cambridge University Press. https://doi.org/10.1017/CBO9780511778568

Kornerup, P. and Muller, J.M. (2006). Choosing starting values for certain Newton–Raphson iterations. Theoretical Computer Science 351(1), 101-110. https://doi.org/10.1016/j.tcs.2005.09.056

Langelandsvik, L.I., Kunkel, G.J. and Smits, A.J. (2008). Flow in a commercial steel pipe. Journal of Fluid Mechanics 595, 323-339. https://doi.org/10.1017/S0022112007009305

LaViolette M. (2017). On the history, science, and technology included in the Moody diagram. Journal of Fluids Engineering ASME 139(3), 030801-030801-21. https://dx.doi.org/10.1115/1.4035116

Mező, I. and Baricz, Á. (2017). On the generalization of the Lambert W function. Transactions of the American Mathematical Society 369(11), 7917-7934. https://doi.org/10.1090/tran/6911
31